\documentclass[conference]{IEEEtran}
\ifCLASSINFOpdf
\else
\fi
\usepackage[cmex10]{amsmath}
\usepackage{ amssymb }
\usepackage[pdftex]{graphicx}
\usepackage{float}
\usepackage{multicol}
\usepackage{lipsum}
\usepackage{ifpdf}
\usepackage{multirow}
\usepackage{epstopdf}
\usepackage{color}
\usepackage{cite}
\usepackage{graphicx}
\usepackage{soul}
\usepackage{balance}
\usepackage{threeparttable}
\usepackage{wasysym}
\usepackage{url}
\usepackage[normalsize]{subfigure}
\usepackage{changepage} 
\usepackage{enumitem}
\usepackage{dblfloatfix}    
\usepackage{afterpage}
\usepackage{changepage} 
\usepackage[hidelinks]{hyperref}
\usepackage{accents}

\usepackage{algorithmicx}
\usepackage{algpseudocode}

\usepackage{fancyhdr}

{}
\begin{document}
%
\IEEEpeerreviewmaketitle

\title{Control and Simulation of a Grid-Forming Inverter for Hybrid PV-Battery Plants in Power System Black Start}
%
%
%


\author{\IEEEauthorblockN{Quan Nguyen, Mallikarjuna R. Vallem,
		Bharat Vyakaranam, Ahmad Tbaileh, Xinda Ke, Nader Samaan\\
		\IEEEauthorblockA{Pacific Northwest National Laboratory, WA}
} }

\maketitle
\thispagestyle{fancy}


\begin{abstract}
	Power system restoration is an important part of system planning. Power utilities are required to maintain black start capable generators that can energize the transmission system and provide cranking power to non-blackstart capable generators. Traditionally, hydro and diesel units are used as black start capable generators. With the increased penetration of bulk size solar farms, inverter based generation can play an important role in faster and parallel black start thus ensuring system can be brought back into service without the conventional delays that can be expected with limited black start generators. Inverter-based photovoltaic (PV) power plants have advantages that are suitable for black start. This paper proposes the modeling, control, and simulation of a grid-forming inverter-based PV-battery power plant that can be used as a black start unit. The inverter control includes both primary and secondary control loops to imitate the control of a conventional synchronous machine. The proposed approach is verified using a test system modified from the IEEE 9-bus system in the time-domain electromagnetic transient simulation tool PSCAD. The simulation results shows voltage and frequency stability during a multi-step black-start and network energization process.
	
\end{abstract}


%

\section{Introduction}
\IEEEPARstart{B}{lack} start (BS) is a process of restoring a power system following a major collapse or a system-wide blackout. This process relies on one or multiple generation units that are able to start without the support from the main power system. Such units are also called BS units. 

The characteristics of an ideal BS unit include small start-up power requirement, fast ramping rate to quickly reach the rated power, large real and reactive power capacity to meet any needs during the black start process, and ability to stabilize system voltage and frequency due to the subsequent energizations of the remaining non-black-start (NBS) units and transmission lines as well as load pickup \cite{Hydrowires_1}. In practice, a BS unit can be a hydro, natural gas or diesel units \cite{FERC_1}. 

Regarding renewable energy sources (RES), solar energy becomes more and more common as the photovoltaic (PV) penetration in distribution feeders has been increasing rapidly in recent years. On the other hand, there are few current and planned large-scale PV power plants that are integrated directly to the transmission network. In \cite{Sandia_1}, general procedures for interconnection of large-scale PV plants and technical requirements for plant performance are discussed. In \cite{NREL_1}, the authors develop a demonstration concept and test plan to show how different power control modes can help an inverter-based PV power plant to provides a wide range of ancillary services. A summary of operational benefits and issues for a large-scale PV power plant including power quality, power control, protection, balancing and reliability under different loading conditions are presented in \cite{Rakhshani_1}.

Considering the advantageous properties such as fast ramping rates, power ratings of several hundreds MW, and ability to coordinate with energy storages (ES), recent large-scale inverter-based PV power plants can be considered as promising BS resources. However, to the best knowledge of the authors, little focus has been given to explore such potential capability of a PV-ES power plant.

A generation system needs several characteristics to be considered as a BS unit and several studies need to be performed to assess them. These studies include power plant modeling, grid-forming inverter control, effective coordination between PV and ES during a black start process, voltage and frequency stability, and sizing of the PV and ES. Note that similar works might exist at distribution voltage level; however, the focus of this paper is on the transmission voltage level. In addition, case studies using multiple test systems with different topologies and parameters are required to thoroughly verify the modeling and control design of the grid-forming inverters. Average (positive sequence) models for inverters do not capture all the characteristics and limitations of inverters. A validation should be carried out using an electromagnetic transient (EMTP) simulation tool and a high-fidelity inverter model to capture detailed dynamic responses.

In this paper, modeling, grid-forming control, conceptual design, and detailed simulation of an inverter-based PV-ES power plant that can be used as a reliable BS resource during a black start are demonstrated. The paper is organized as follows. Section II briefly describes current black-start practice using a conventional synchronous generator. Section III presents the proposed modeling, control, and operation of a PV-ES plant during a black start. In Section IV, the simulation results of a black-start process in a test system in the time-domain EMTP simulation tool PSCAD are discussed. Final conclusions and remarks are given in Section V.

\section{Black-Start Process and State of The Art}
Current practice of a black start process includes several steps\cite{Adibi_1, Hydrowires_1}. First, one or multiple black-start (BS) units such as hydro power plants start their own power without support from the grid. Next, main transmission lines and transformers are energized to form the backbone of the system. A small potion of load might be picked up during this process to assure voltage stability or handle transformer inrush current. Finally, non-black-start (NBS) units and the remaining loads are gradually picked up by energizing the rest of transmission lines while maintaining the voltage and frequency of the system.

\begin{figure}[t!]
	\centering
	\includegraphics[width = 1\columnwidth] {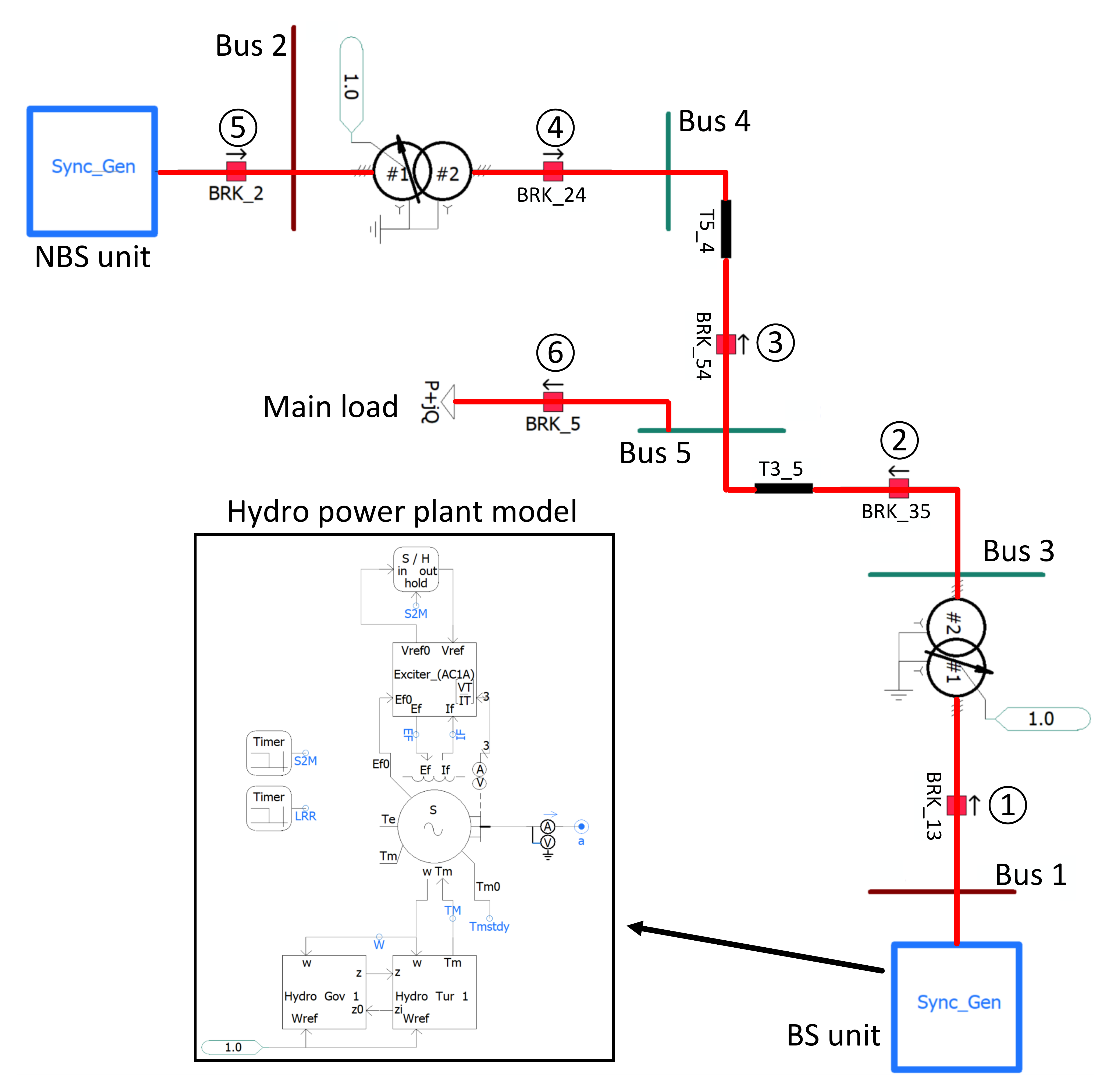}	
	\vspace{-0.7cm}	
	\caption{A test system for black-start demonstration using a conventional hydro generator as a BS unit.}
	\label{fig:IEEE_Conventional}
\end{figure}

Fig. \ref{fig:IEEE_Conventional} shows an example of the black-start process using a 5-bus system. The BS hydro power plant at Bus 1 and circuit breakers to pick up a NBS unit at Bus 2 and the main load at Bus 5. The simulation is conducted in the time-domain electromagnetic transient tool PSCAD \cite{PSCAD}. The circles with numbers show the sequential order of the energizing and picking up transformers, transmission lines, generators, and load. 

\begin{figure}[t!]
	\centering
	\includegraphics[width = 1\columnwidth] {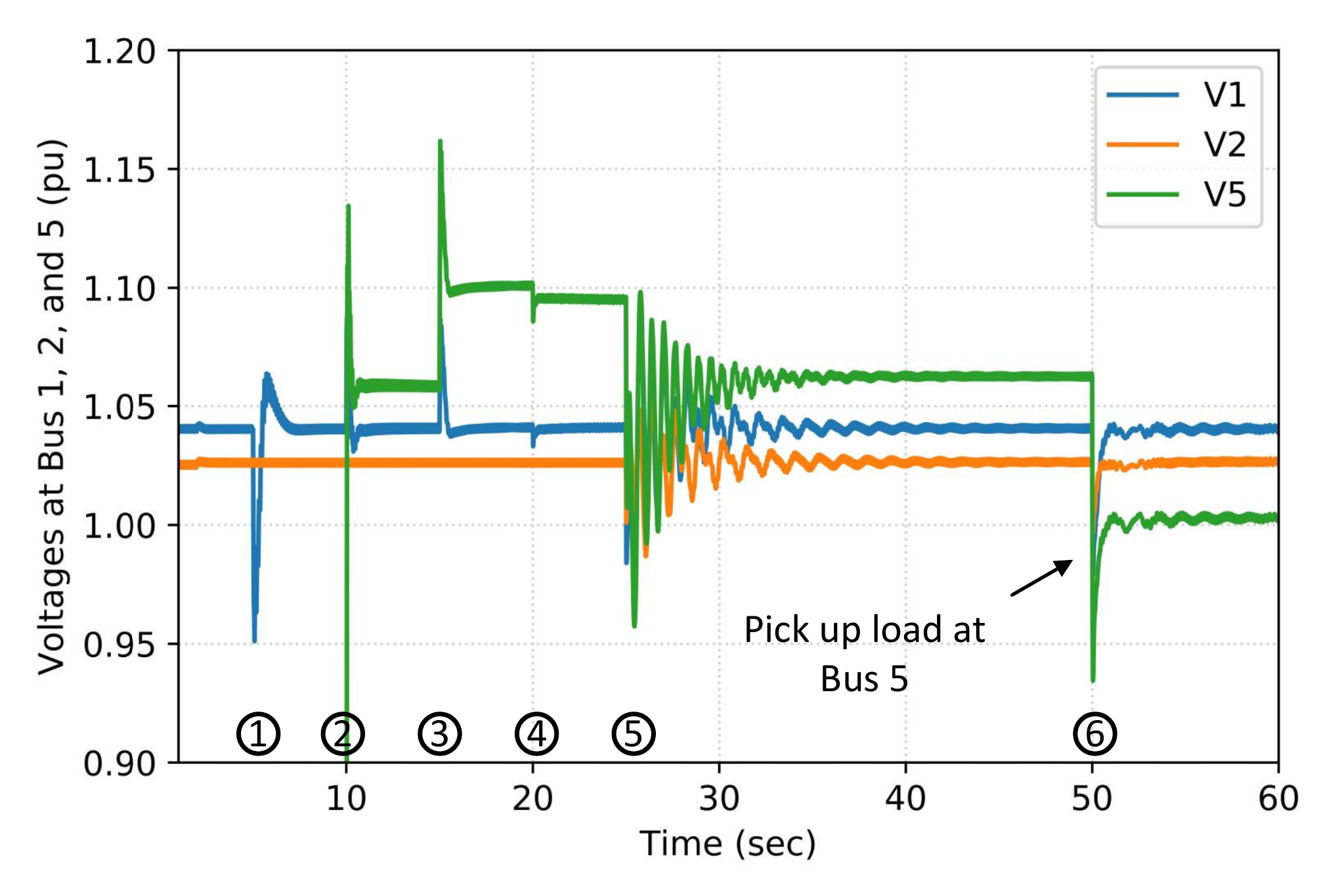}	
	\vspace{-0.8cm}		
	\caption{Voltage profiles at Buses 1, 2, and 5 during the black-start process.}
	\label{fig:IEEE_Conventional_Vbus}
\end{figure}

\begin{figure}[t!]
	\centering
	\includegraphics[width = 1\columnwidth] {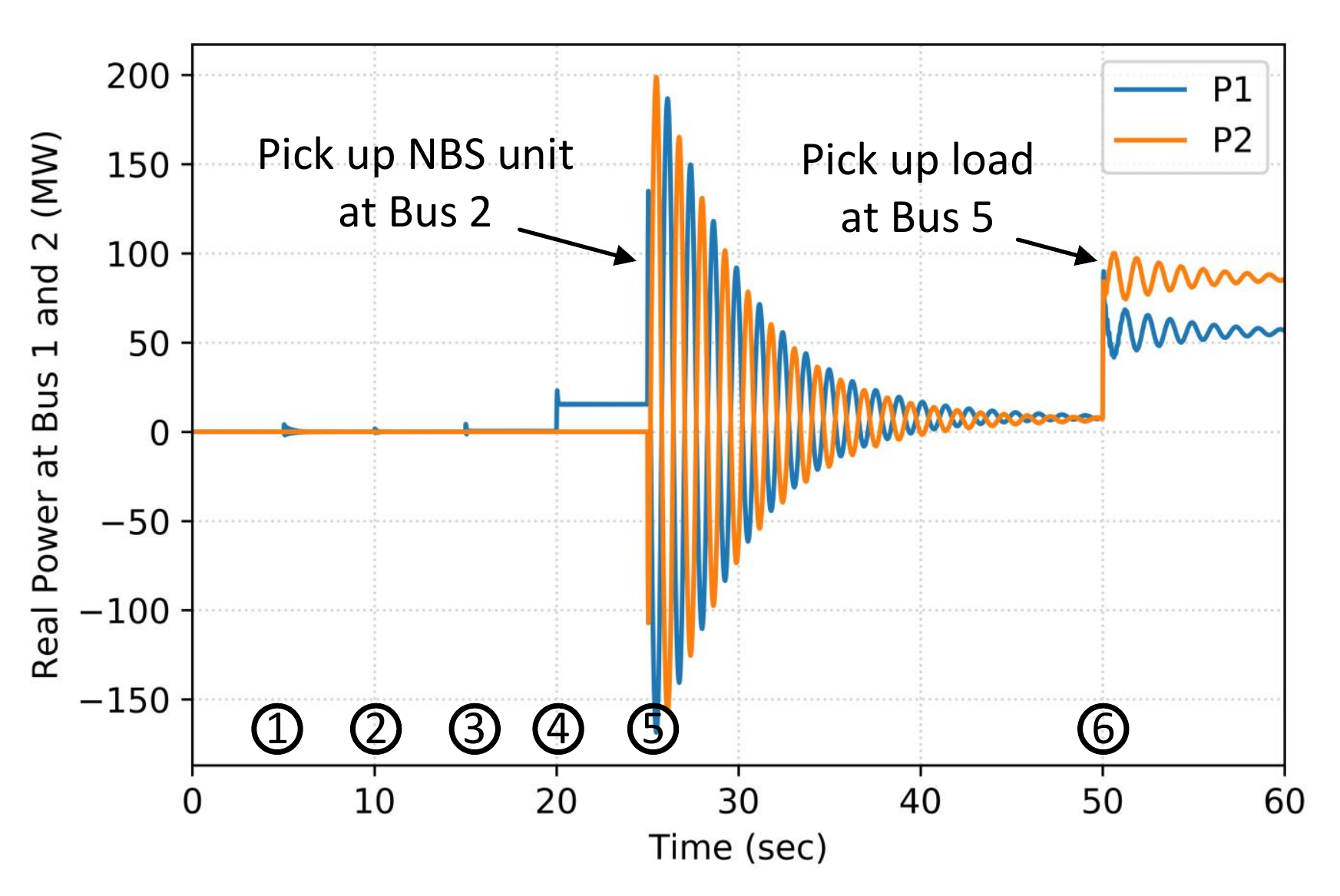}	
	\vspace{-0.8cm}	
	\caption{Real power generation from BS unit at Bus 1 and NBS unit at Bus 2.}
	\label{fig:IEEE_Conventional_P}
\end{figure}

Figs \ref{fig:IEEE_Conventional_Vbus} and \ref{fig:IEEE_Conventional_P} show the resulting voltage at Buses 1, 2, and 5 as well as the power generation of the BS and NBS units at Buses 1 and 2, respectively, during the black-start process. As expected, these simulation results show stable responses after each switching event of circuit breakers. 

\section{Modeling, Control, and Operation of a Grid-Forming Inverter-Based PV-ES plant for Black Start}\label{Sec: Inverter Modeling and Control}
	This section presents the modeling and its grid-forming control of an inverter-based PV-ES power plant that is capable of being a BS unit for a black-start process.
	
\begin{figure*}[t!]
		\centering
		\includegraphics[width = 1.7\columnwidth] {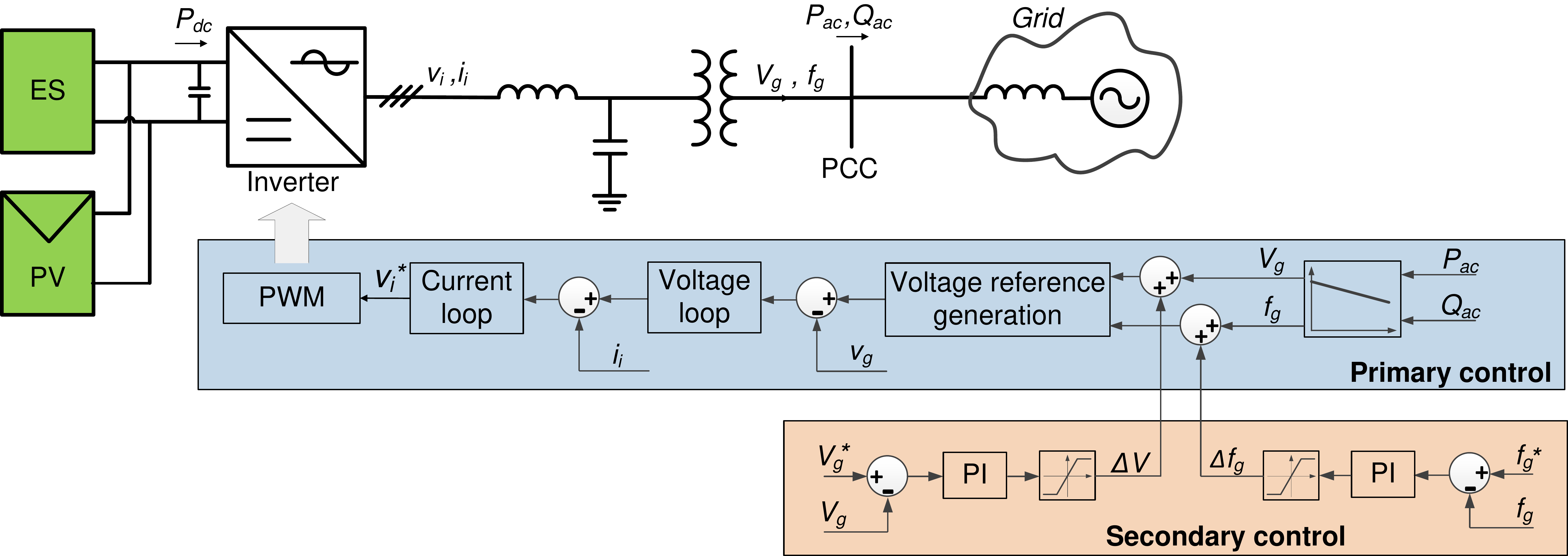}	
		\caption{The grid-forming control of an inverter-based PV-ES power plant.}
		\label{fig:Inverter_Control}
	\end{figure*}

	The physical structure and grid-forming control of the PV-ES power plant is shown in Fig. \ref{fig:Inverter_Control}. The DC side of the inverter system consists of PV arrays and energy storages connected in parallel. The AC side consists of switching components that form an H-bridge, a low-pass LC filter, and a step-up transformer.
	
	\subsection{Grid-Forming Inverter Control}
	As shown in Fig. \ref{fig:Inverter_Control}, the control of a grid-forming inverter-based PV-ES power plant includes both primary and secondary control loops, which imitates the control of conventional synchronous machines. This control strategy is adapted from the inverter control for microgrids at distribution level \cite{Shafiee_1}.
	
	The primary control loop modifies the voltage magnitude and frequency at the point of common coupling (PCC) based on a pre-defined droop curve and the measured real and reactive power $P_{ac}$ and $Q_{ac}$ injected to the grid. The resulting voltage magnitude and frequency, both are deviated from the rated values, are used to generate a three-phase sinusoidal voltage signal. Such a signal is the input of classical nested voltage- and current- loops, designed in the synchronously rotating reference frame $dq$ \cite{Teodorescu_1}, to create the inverter reference voltage $v_i^*$. The Pulse-Width-Modulation (PWM) block takes $v_i^*$ and generates switching signals for the switching devices of the inverter. 
	
	The secondary control loop is responsible for restore the voltage and frequency back to rated values. This goal is achieved by adding quantities $\Delta V_g$ and $\Delta f_g$, which are the output of the secondary PI controllers, to the output voltage $V_g$ and frequency $f_g$ of the primary droop controllers.

	\subsection{Modeling and Control of PV and ES}
\begin{figure}[t!]
		\centering
		\includegraphics[width = 0.8\columnwidth] {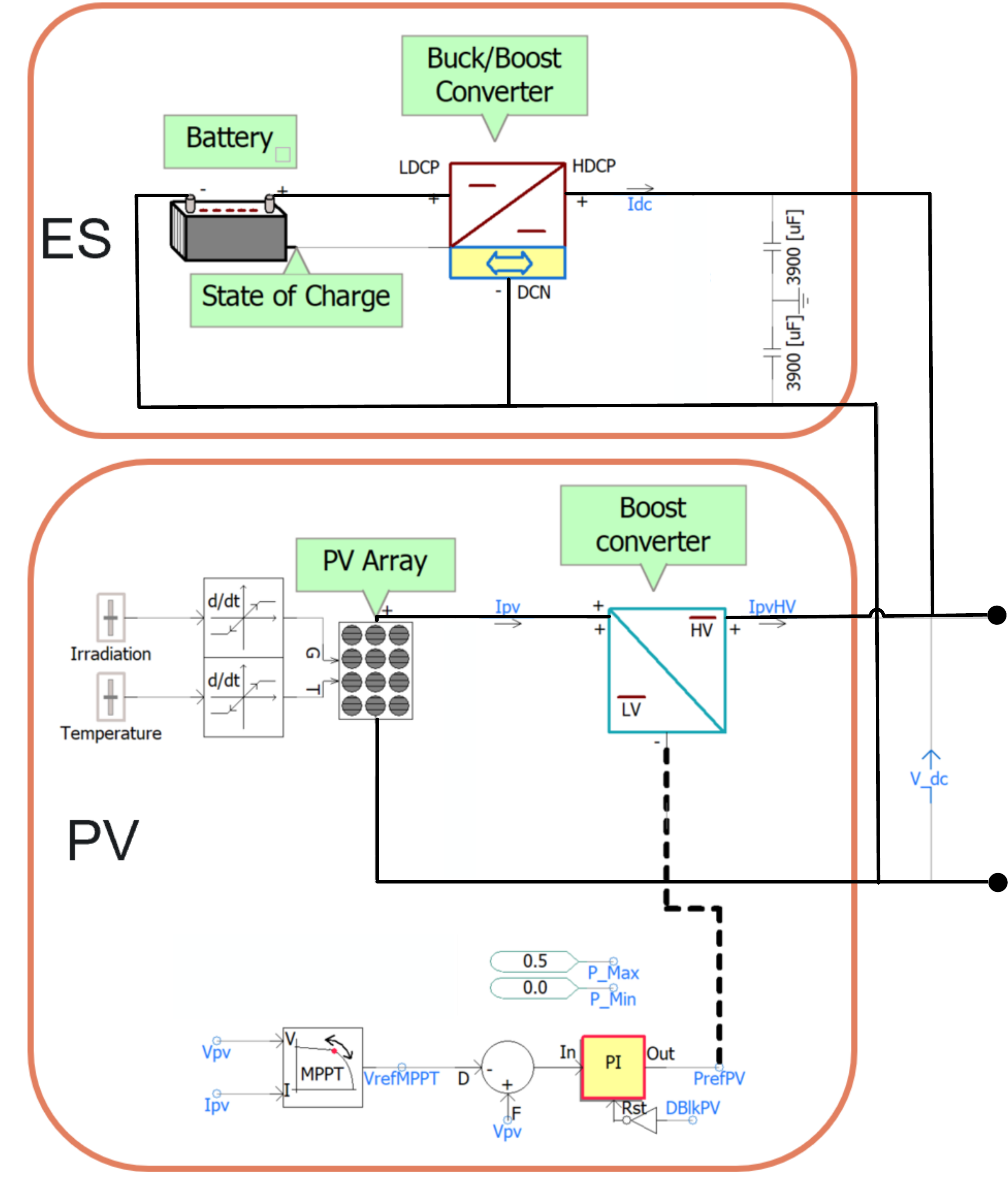}
		\vspace{-0.2cm}	
		\caption{The modeling of PV array and ES at the DC side of an inverter-based PV-ES power plant \cite{PSCAD_2}.}
		\label{fig:DC_side}
	\end{figure}

	Fig. \ref*{fig:DC_side} shows the modeling of the DC side of an inverter, which is the combination of a PV array and an ES. The models of the PV array and ES are leveraged from \cite{PSCAD_2}. Due to the intermittence of solar, the ES is required in the PV-ES power plant model to guarantee a successful black start. 
	
	The input of the PV inverter includes the solar irradiation and temperature. A boost converter is used to boost the output voltage of the PV array to a sufficient inverter input voltage for achieving the desired AC voltage magnitude at the AC side. On the other hand, the ES model includes a detailed battery model and a conventional buck-boost converter. The DC voltage controller of the buck-boost converter is responsible for keeping the voltage across the DC-link capacitors, which are shared by the PV array and ES, at a constant value.
	\begin{figure*}[t!]
		\centering
		\includegraphics[width = 1.7\columnwidth] {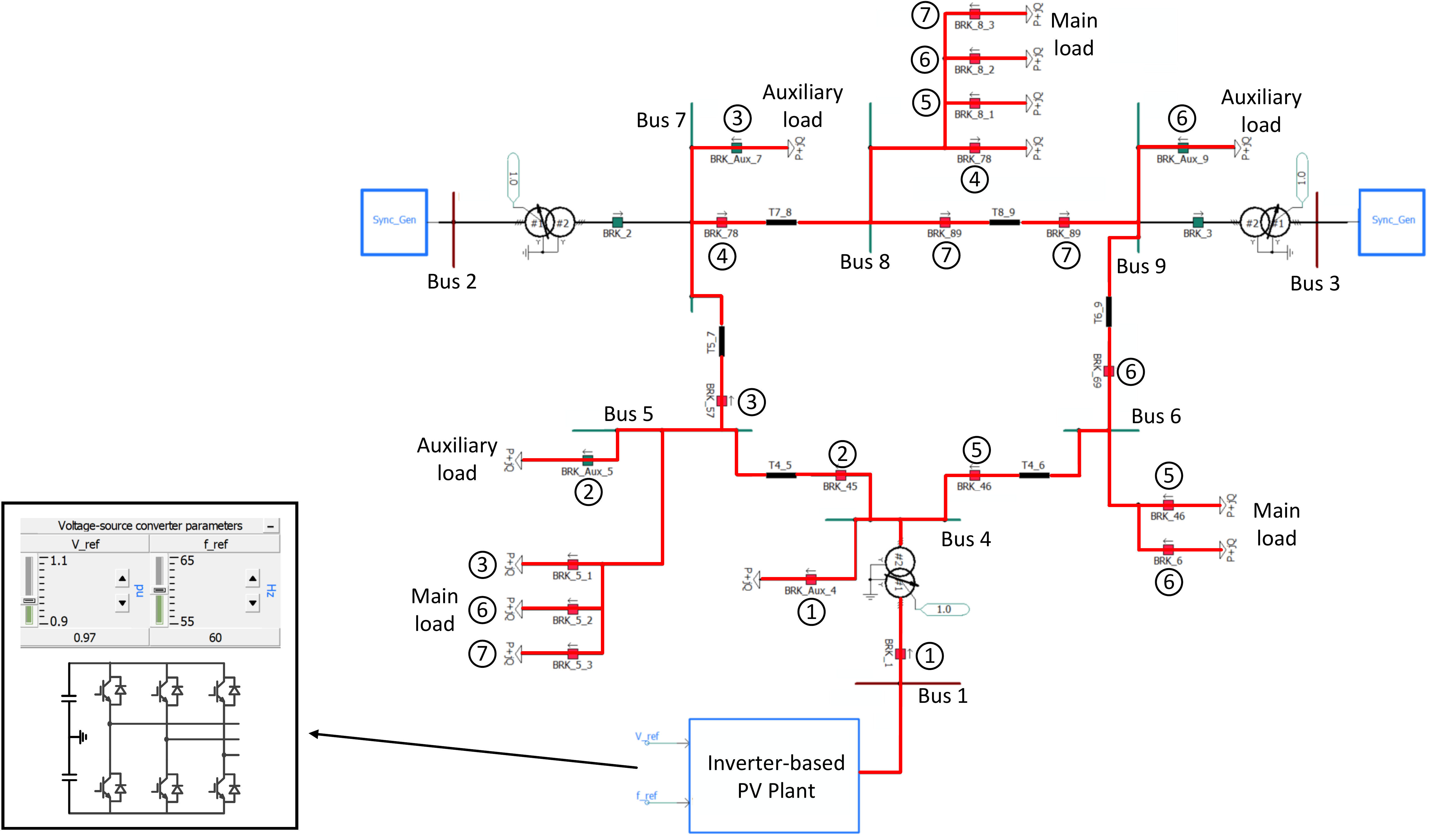}	
		\caption{Test system for simulating back-bone energizing using a PV-ES power plant at Bus 1 as a BS unit.}
		\label{fig:IEEE_9bus}
	\end{figure*}

	\subsection{Operational Concept during a Black-start Process}
	During the entire black-start process, the voltage magnitude and frequency at the PCC are maintained at the rated values. On the other hand, the supplied power $P_{ac}$ and $Q_{ac}$ at the PCC vary with each switching event  of circuit breakers to energize transformers and transmission lines as well as pick up NBS units and loads. The real power $P_{ac}$ required by the black start process entirely depends on the support from the PV array and ES at the DC side. Therefore, it is important to size the PV and ES accordingly.
	
	At the DC side, as input DC-link voltage of the inverter is kept constant during the entire black start, the controls of the DC and AC sides are decoupled. After each switching event of circuit breakers during the black start process, the PV array and ES varies their output real power to meet the real power requirement at the AC side. In this work, the ES is assigned as the primary power supply to the AC side as long as its state of charge is positive. If the real power requirement from AC side exceeds the upper limit of discharging power of the ES, the remaining power required is provided from the PV array.

	\section{Case Study}
	This section shows the time-domain electromagnetic transient simulation in PSCAD to demonstrate the black-start capability of the PV-ES power plant described in Section \ref{Sec: Inverter Modeling and Control}. More specifically, the simulation focuses on the backbone energization of the system, i.e. energizing main transmission lines and transformers as well as picking up loads. The metrics used to validate the efficacy of the grid-forming control include two folds. First, the voltage and frequency must be stable during the entire backbone-energizing process. Second, an optimal sequence of energizing transmission lines and transformers as well as optimal amounts of load from a separated optimization problem are used as the input of the simulation. Therefore, another metric to verify the grid-forming control is the similarity between the steady-state voltages from the simulation result and the voltages from  the numerical optimization solution.
	\begin{table}[t!]
		\renewcommand{\arraystretch}{1.3}
		\setlength{\tabcolsep}{0.6em}
		\caption{Total amount of picked-up load at each bus.} 
		\label{tab:Pickup_Load}
		\centering
		\begin{tabular}[h]{|c|c|c|c|c|c|c|c|}
			\hline
			Step 	& Time (s)& $P_5^{load}$ & $Q_5^{load}$ & $P_6^{load}$ & $Q_6^{load}$ & $P_8^{load}$ & $Q_8^{load}$	\\
			\hline									 						
			1 	&	0.0		& 0.0 	& 0.0   & 0.0	 & 0.0    & 0.0   & 0.0 	\\ 
			2 	&	1.0		& 0.0 	& 0.0   & 0.0	 & 0.0    & 0.0   & 0.0 	\\ 
			3 	&	2.5		& 13.4	& 5.3   & 0.0	 & 0.0    & 0.0   & 0.0 	\\ 
			4 	&	5.0		& 13.4	& 5.3   & 0.0	 & 0.0    & 28.8  & 10.1	\\ 
			5 	&	7.5 	& 13.4	& 5.3   & 4.0  & 1.3    & 33.4  & 11.7	\\ 
			6 	&	10.0 	& 26.1	& 10.4  & 59.8 & 19.9   & 34.3  & 12.0	\\
			7 	&	18.0 	& 30.1	& 12.0  & 59.8 & 19.9   & 65.7  & 23.0	\\
			\hline
		\end{tabular}
	\end{table}

	\subsection{System Description}	
	The test system used in this section is the IEEE 9-bus system, as shown in Fig. \ref{fig:IEEE_9bus}. The BS start unit is assumed to be located at Bus 1. However, instead of using a conventional hydro plant as the BS unit as shown in Fig. \ref{fig:IEEE_Conventional}, an inverter-based PV-ES power plant with grid-forming capability is deployed. The ratings of the PV and ES are assumed to be sufficient to meet the power requirement during the entire backbone-energizing process. The optimal switching sequence of the circuit breakers to energize transmission lines are represented by the circles with numbers. The picked up loads at Buses 5, 6, and 8 at each step are shown in Table \ref{tab:Pickup_Load}. 
	
	During the simulation, it is observed that the energization of transformers and transmission lines might negatively affect system stability. The former causes inrush current that results in undervoltage violations while the latter results in overvoltage violations during a low load condition. Therefore, small amount of auxiliary loads might be necessary at the secondary of a transformer and the end terminal of a transmission line to guarantee voltage stability. For example, an auxiliary load is added at Bus 4, which is the secondary side of the transformer between Buses 1 and 4. In addition, auxiliary loads are added at Buses 5, 7, and 9 when energizing the lines 4-5, 5-7, and 6-9, respectively. At the end of the simulation when the main loads are picked up, these auxiliary loads are disconnected. 
	\begin{figure}[t!]
		\centering
		\includegraphics[width = 1\columnwidth] {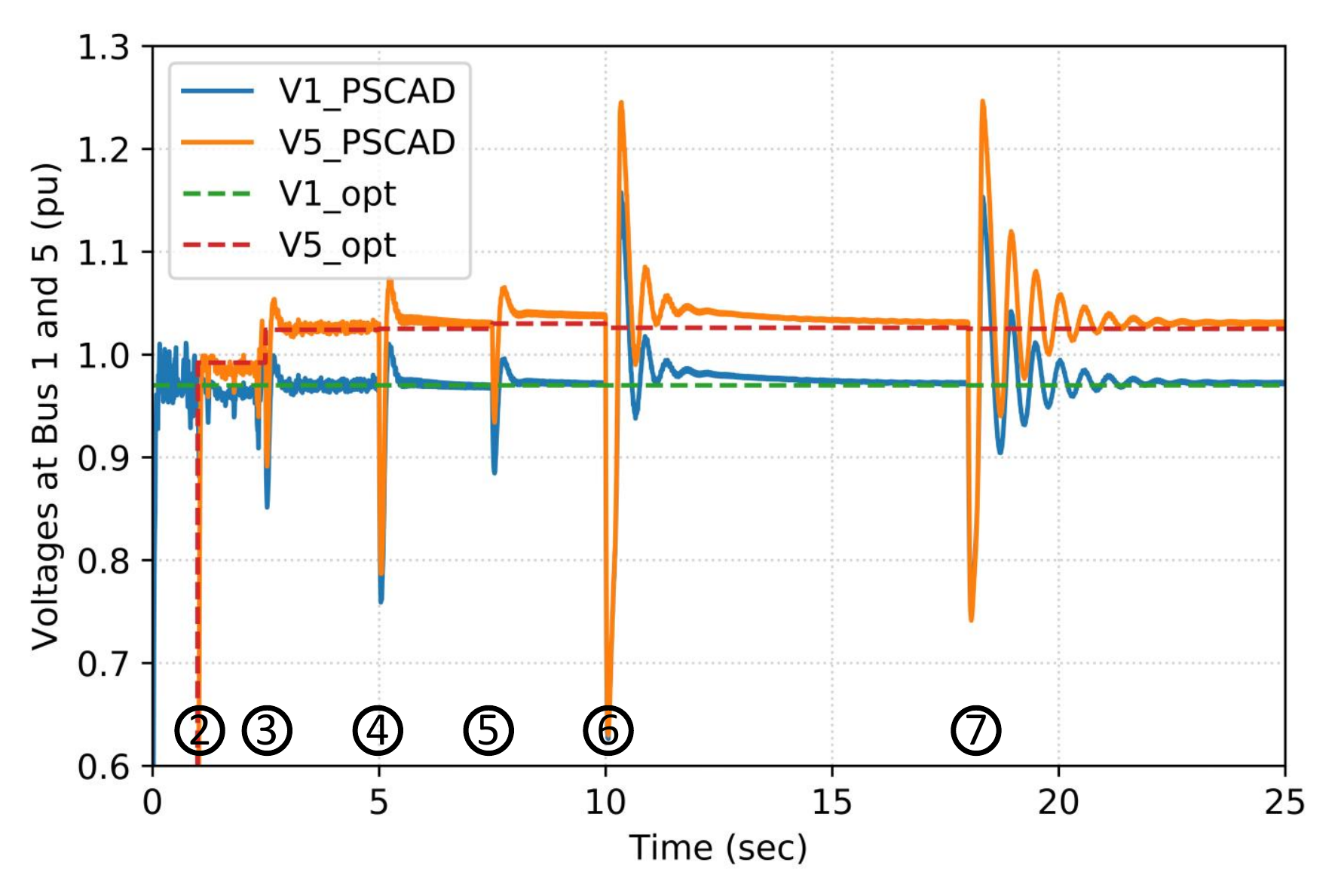}	
		\vspace{-0.8cm}
		\caption{Bus voltages during the backbone energizing process.}
		\label{fig:Bus_voltages}
	\end{figure}
	\begin{figure}[t!]
		\centering
		\includegraphics[width = 1\columnwidth] {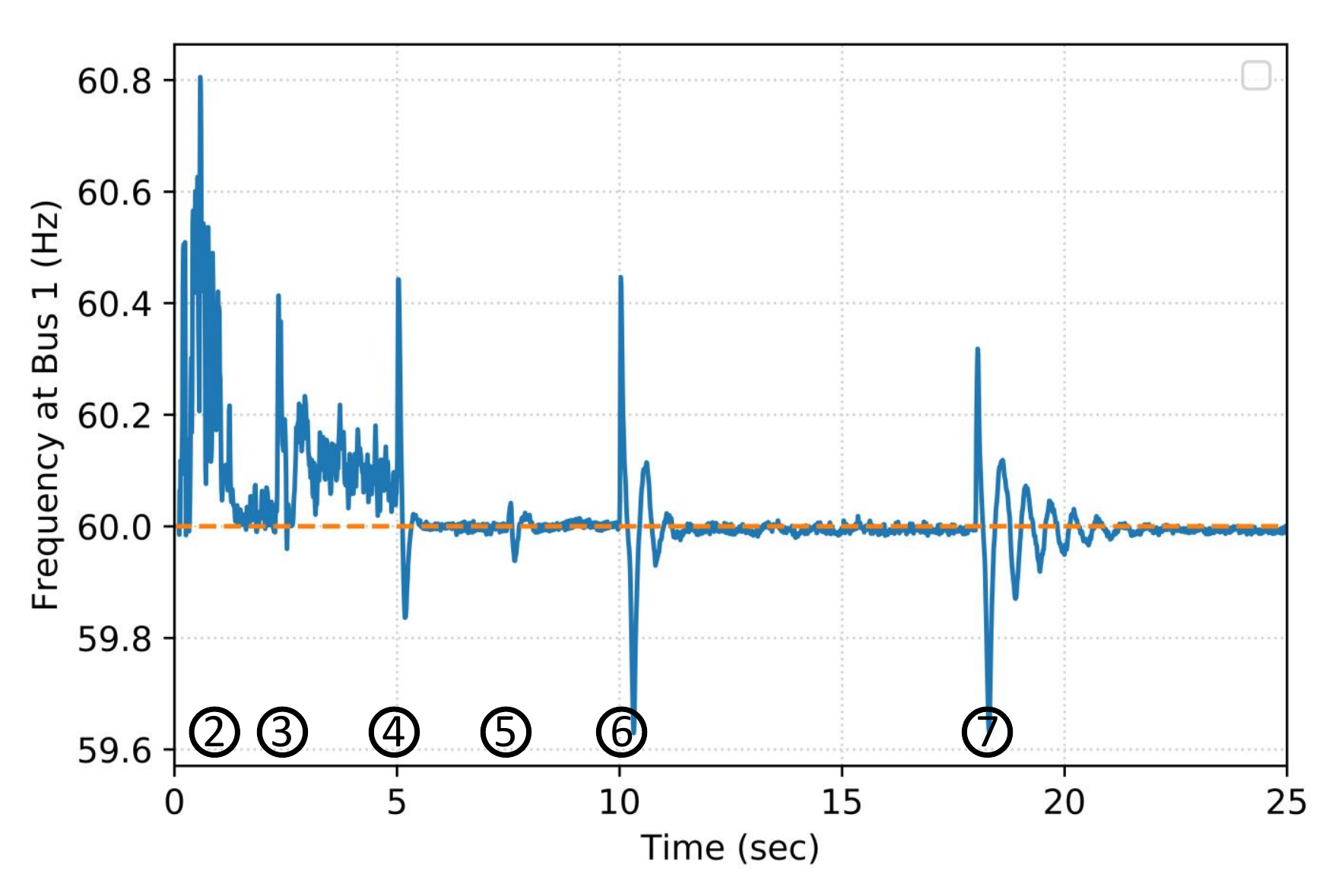}	
		\vspace{-0.8cm}
		\caption{Frequency measured at Bus 1 during the backbone energizing process.}
		\label{fig:Frequency}
	\end{figure}
	\begin{figure}[t!]
		\centering
		\vspace{-0.3cm}
		\includegraphics[width = 1\columnwidth] {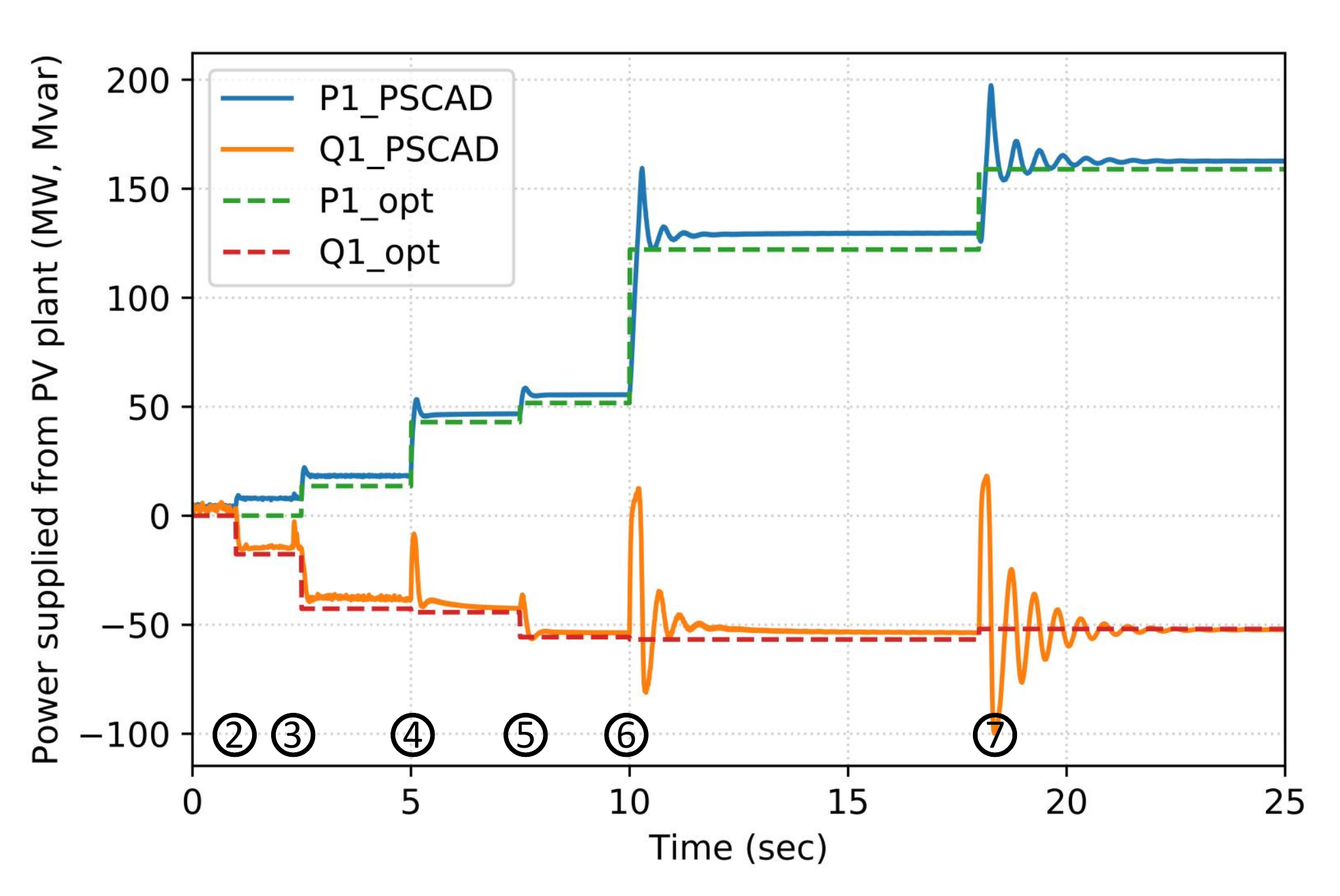}	
		\vspace{-0.8cm}
		\caption{Power supplied by the PV-ES power plant at Bus 1 during the backbone energizing process.}
		\label{fig:Gen_power}
	\end{figure}
	\begin{figure}[t!]
		\centering
			\vspace{-0.3cm}
		\includegraphics[width = 1\columnwidth] {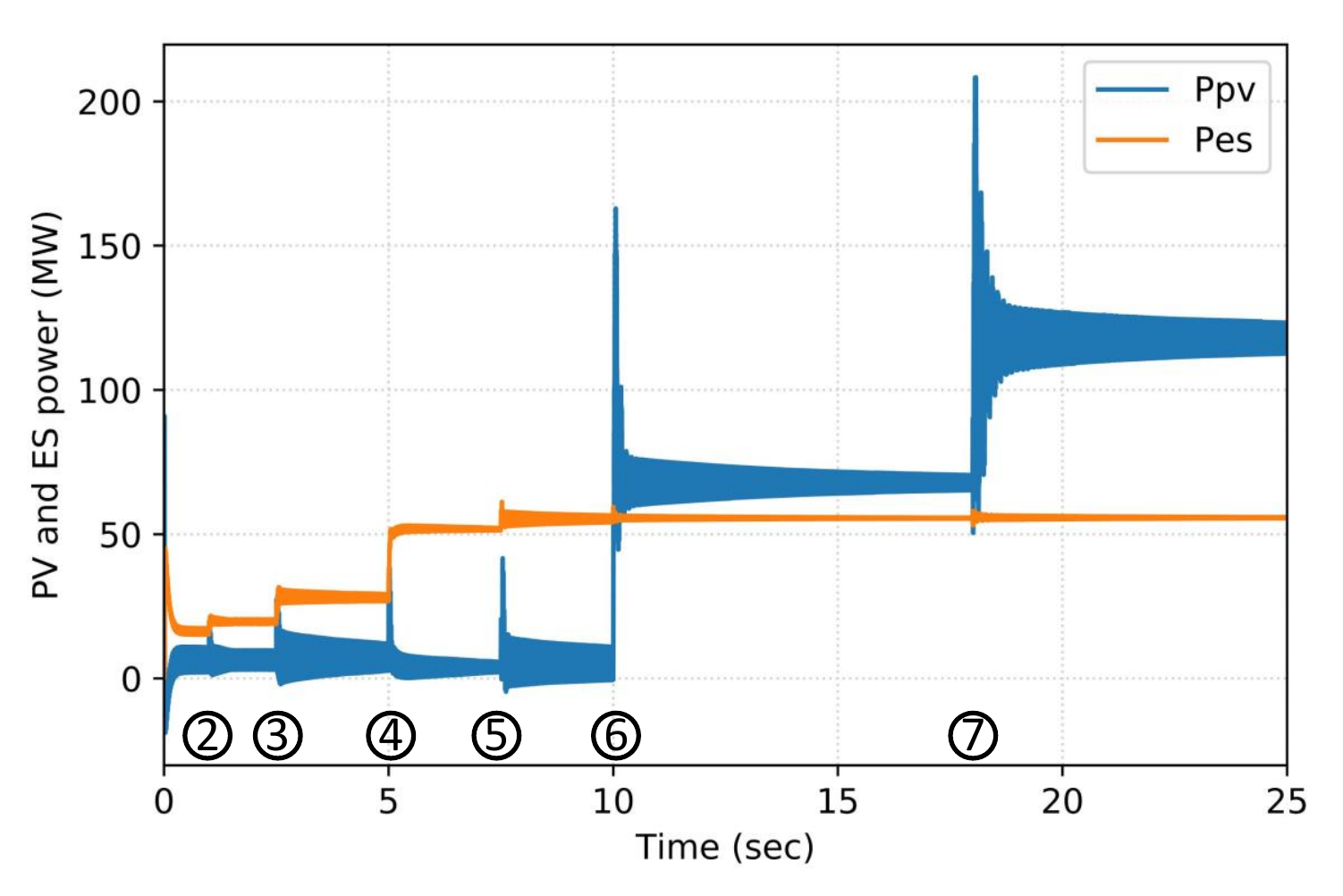}
		\vspace{-0.8cm}	
		\caption{Power supplied by the PV array and ES at the DC side.}
		\label{fig:Ppv_Pes}
	\end{figure}

	\subsection{Simulation result}
	Fig. \ref{fig:Bus_voltages} shows the voltages at Buses 1 and 5 from the PSCAD simulation result and the numerical optimization solution. It can be seen that the simulated voltages are stable after all switching events, and their stead-state values closely match the numerical optimization solution. 
	
	Fig. \ref{fig:Frequency} shows the frequency at Bus 1 where the PV-ES power plant is located. The is an acceptable noise in the frequency profile during the initial no-load condition from 0 to 5 seconds. However, after the main loads are picked up, the resulting frequency is also stable and its steady-state value closely matches the rated value of 60 Hz. 
	
	Fig. \ref{fig:Gen_power} shows the real and reactive power supplied from the PV-ES power plant from the simulation result and numerical optimization solution. It can be seen that the simulation result and numerical solution are also similar to each other.
	
	Fig. \ref{fig:Ppv_Pes} shows the power supplied from the PV array and ES at the DC side of the PV-ES power plant. From the beginning to 5 seconds, the ES is the main supply resource as the required power from the AC side is less than the discharging power limit of the ES. However, after 5 seconds, the power supplied from the ES remains constant while the remaining power required from the AC side is provided by the PV array.

\section{Conclusion}
	In this paper, AC- and DC-side modeling, a grid-forming control, and an operational concept for an inverter-based PV-ES power plant with ES are proposed. A time-domain EMTP simulation for the IEEE 9-bus system is used to evaluate the efficacy of the proposed approach. The simulation results show completed voltage and frequency stability during the entire black start process as well as high control accuracy when comparing the simulation result and a numerical input. Based on these results, a PV-ES power plant using the proposed work can be successfully and reliably used as a BS unit for a black start in a transmission system.

\section{Acknowledgment}
	This material is based upon work supported by the U.S. Department of Energy's Office of Energy Efficiency and Renewable Energy (EERE) under the Solar Energy
	Technologies Office Award Number DE-EE0008770.

\bibliography{QuanNguyen_References}
\bibliographystyle{IEEEtran}

\end{document}